# Measuring the impact of an instructional laboratory on the learning of introductory physics

Carl Wieman

*Department of Physics, Stanford University, Stanford CA 94305*

*Graduate School of Education, Stanford University, Stanford CA 94305*

N.G. Holmes

*Department of Physics, Stanford University, Stanford CA 94305*

(Dated: May 22, 2015)

Abstract

We have analyzed the impact of taking an associated lab course on the scores on final exam questions in two large introductory physics courses. Approximately a third of the students who completed each course also took an accompanying instructional lab course. The lab courses were fairly conventional, although they focused on supporting the mastery of a subset of the introductory physics topics covered in the associated course. Performance between students who did and did not take the lab course was compared using final exam questions from the associated courses that related to concepts from the lab courses. The population of students who took the lab in each case was somewhat different from those who did not enroll in the lab course in terms of background and major. Those differences were taken into account by normalizing their performance on the lab-related questions with scores on the exam questions that did not involve material covered in the lab. When normalized in this way, the average score on lab-related questions of the students who took the lab, in both courses, was within 1% of the score of students who did not, with an uncertainty of 2%. This result raises questions as to the effectiveness of labs at supporting mastery of physics content.

## I. Introduction

Instructional labs are a major part of undergraduate physics education, particularly at the introductory level. They involve substantial instructional resources due to the need for dedicated space and equipment and a relatively low student-to-instructor ratio. As such, it is important to measure their actual educational contribution to ensure these resources are being used widely.

To many physicists, instructional labs are considered essential for a proper physics education, and they are required for AP physics courses and as part of many state K-12 science standards. There has been little research, however, on what learning such laboratories actually achieve. America's Lab Report[1] reviewed the research on science instructional laboratories and found that there was little research in general. What there was provided little evidence of effectiveness, particularly for stand-alone laboratory courses, though the research is very thin and most of it is relatively old.[2-5] We are distinguishing lab courses, held in different locations and times from the introductory course lectures, but either receiving a separate course grade or a grade that is a composite of the two parts of the course, from the studio physics approach.[6-10] In studio physics courses there is complete integration of all aspects of the course, which includes a number of small experiments that are carried out in the classroom. There is solid evidence of the learning achieved with such studio physics courses, but, because of their integrated nature, it is impossible, and arguably meaningless, to try and isolate what component of that is provided by the experimental portion. Given the added resources and novel teaching methods needed for such studio courses, it is unlikely most institutions will be adopting them in the near future so it remains important to examine the learning that is produced in stand-alone lab courses.

As noted in America's Lab Report, the biggest challenge with regard to research on the educational effectiveness of labs is the lack of clarity with regard to the educational goals.[1] The instructional lab community often emphasizes using experiments to present particular physics concepts, especially to reinforce material from lecture. Learning physics concepts through lab experiments might resemble learning through demonstrations. Although not all demonstrations improve learning, interactive lecture demonstrations, where students predict the outcome of a demonstration prior to seeing it and then reflect on the results, have been shown to improve students' conceptual understanding.[11-13] It is possible that seeing physical phenomena when carrying out a hands-on experiment might lead to similar gains in physics knowledge, especially if students are asked to make predictions and reflect on results in a similar way.

While many instructional labs, including the ones studied in this paper, present explicit goals for content mastery, there are often many additional goals of labs that are intended, though often only implicitly addressed. If one judges the goals by what is required and counted in grading, it is customary to have a lab course also include the following goals: learning to keep a lab notebook; carrying out proper analysis of experimental uncertainty; learning to work with complex equipment; learning to take data in the optimum manner; learning to correctly interpret the comparison of uncertain data with mathematical models; learning to write up an experiment; and generally learning the process of science as an experimental activity. This diversity of goals was reflected in the American Association of Physics Teachers' (AAPT) published set of goals for introductory physics labs.[14] These were summarized as: the art of experimentation; experimental and analytical skills; conceptual learning; understanding the basis of knowledge in physics; and developing collaborative learning skills.

In recent years, there has been growing discipline-based education research (DBER) interest in instructional labs. Much of the research has explored student attitudes about labs,[15-19] student understanding of uncertainty, measurement, and data analysis,[20-24] and student development of scientific reasoning and experimentation skills.[25-28] The learning of the physics content, including the understanding and application of concepts, however remains a common goal of physics labs. Presumably, this is because many physicists see experimental research, both currently and in the past, as how physics knowledge is established, and so it is believed that replicating the discovery process will support student learning of the knowledge. The effectiveness of such learning remains poorly evaluated, however. Since students are typically exposed to this physics content through lectures, labs, homework, and recitation sections, isolating the contribution from labs is challenging. Here we took advantage of some particularly favorable circumstances to measure the impact of learning physics content from a lab course that was relatively standard in its structure, but optimized in a number of respects.

In this study we examined two stand-alone introductory physics lab courses that were both closely coordinated with the standard first year mechanics and electricity and magnetism courses for students planning to major in engineering or most of the sciences. The goals, design, and structure of these lab courses are unusually well defined as detailed below, and focused on increased understanding of the concepts covered in the mechanics and E & M courses. We examined the impact of taking the two lab courses on students' performance on the exam questions in the two respective lecture courses.

## II. Setting

The context for this study is the first and second term courses of the introductory calculus-based physics sequence at a large elite institution. The first course is the primary introductory mechanics course, "physics 1," and the second is the primary introductory electricity and magnetism course, "physics 2." More than 500 students take each of these courses per year, most of whom are intending to major in various fields of engineering. About half the physics majors and some other science majors such as chemistry also take these courses, but they make up a small fraction of the enrollment. The physics 1 course covers the standard mechanics topics, from vectors up through rotational motion. The physics 2 course covers topics from electrostatics, simple AC and DC circuits, to Maxwell's equations. The physics 1 course is pre-requisite for physics 2. The textbook for both courses is Knight, Introduction to Physics (Volumes 1 and 4), a widely used textbook. Both instructors used interactive engagement methods in lecture such as peer instruction (supported by clickers) and interactive demonstrations. The courses are also accompanied by a one-hour recitation session, which primarily focuses on solving problems in small groups related to the lecture material, supported by a teaching assistant (TA).

Accompanying both lecture courses is an associated laboratory course, also led by TAs. These are one-credit stand-alone courses with their own grade (pass-fail) taken by approximately one third of the students who take the respective lecture course (physics lab 1: 211 students out of 571 students in physics 1; physics lab 2: 169 students out of 530 students in physics 2). The physics 1 lab is not pre-requisite for the physics 2 lab, so students in either lab may or may not have had previous university lab experiences. The reason for taking the lab is primarily determined by the student's major and career goals. The labs are well coordinated with the lecture course in timing and content. Both lab courses involve nine experiments, each done in a two-hour period per week. They are conventional topics and experimental design, but with goals explicitly focused on improving understanding of a number of key concepts covered in the lecture courses, as shown in the examples below (full list given at end of paper).

"*The goals of this lab are to understand: The vector nature of forces; The interaction of multiple forces in two dimensions.*"

"*Energy is neither created nor destroyed but can be converted from one form to another. In this lab you will explore two different forms of energy, kinetic and potential. (Knight sections 10.2, 10.4, 10.5)*"

"*The purpose of this lab is to: Understand how charges distribute on a conductor; Map electric potentials for various charge configurations and; Develop an intuitive understanding of the relationship between fields and potentials*"

### C. Lab structure

Most of the student lab work takes place during the scheduled lab period, which is a 2-hour block of time each week led by one TA. All TAs take part in a 1-unit "Teaching of Physics" Seminar course prior to their first quarter of teaching. In addition to discussing interactive teaching and learning techniques, the TAs observe other TAs in lab and/or discussion sections, work through grading activities, and teach a lab or discussion section, for which they are observed and receive feedback. The TAs meet weekly with the course instructor, a head TA, and the lab coordinator to work through the lab activities. They also attend lecture regularly and run office hours, which helps them stay on top of the content material. In each lab section, the TAs prepare a short introductory lecture to go over the lab's concepts and equipment. For the remainder of the time, they visit individual groups to answer questions or probe their understanding through Socratic questioning techniques.

About three quarters of the students have their lab class on Wednesday or Thursday, with the remainder on Friday, with timing set to coordinate coverage of topics in the lab with that in the lecture classes. Each lab includes a pre-lab activity, which is completed before students enter the lab. The pre-lab provides information to prepare students for the lab experiment, focusing on introducing students to the equipment they will be using (for example, there is one activity with questions targeted at understanding the sampling technique used by Logger Pro), technical skills they will need (for example, there is one activity with questions targeted at understanding the nature of log-log plots), and the physics concepts they will be using in the experiment. Depending on the experiment, pre-lab activities range from a page of text with one or two probing questions, to several pages of introductory material and lab instructions to be read, to five pages of pre-lab questions to be completed. Many of the pre-lab activities include sequences of questions where students explore the relevant physics concepts to make predictions about what they will encounter in the lab experiment (for example, through sketching curves and graphs or completing simple calculations).

Some of the activities were adapted from the *Tutorials in Introductory Physics*[29] or materials from the University of Illinois at Urbana-Champaign introductory physics labs.[30] The apparatuses are standard (Pasco carts, force and motion sensors, etc.) and involve the standard experiments with this equipment, but with considerable focus on sense making, reflecting the goals of the labs. For example, there are regular, explicit prompts for students to compare their observations to predictions, or to look for discrepancies or surprises in their data and to explain their causes. To allow students to focus on the physics concepts, the experiments in the lab are kept quite simple (without complex equipment) and closely tied to the material presented in the lecture and text. The lab instructions and pre-lab activities often refer to specific associated sections in the text. There is minimal or no error analysis, and the write-ups of results only involve students filling out a worksheet. Logger Pro is often used for data collection to make data analysis more efficient, allowing students to spend more time on the concepts. The lab manuals provide detailed instructions, hints, or additional information to avoid issues or complications. For example, in an elastic collisions experiment in the physics 1 lab, students are instructed to keep the initial speed of the carts relatively low to avoid inelastic collisions. In short, these labs are doing just about everything one might hope for in creating an instructional lab course designed to support the learning of the concepts presented in an associated lecture course.

## III. Methodology

Our goal is to determine the effect of taking the lab course on the performance on the physics 1 and physics 2 final exams. This is arguably a proxy for the impact of the lab on the learning of the respective physics. If the two populations (those that take the lab and those that do not) were equivalent, and all the exam questions related to the material covered in the lab course, we could simply look at the average final exam score for each of the two populations, lab and no-lab. However, this has two problems. First, the two populations are different, they tend to have different majors, with associated differences in preparation and attitudes, and second, the lecture courses, and correspondingly, their final exams, cover significantly more material than what is covered by the lab courses. This second fact allows us to use the scores on the non-lab-related exam questions to normalize for the differences in the two populations, and then compare their performance on final exam questions that involved concepts covered in the lab course.

The two courses have different instructors (neither of which were the authors) who have different styles, but in both cases the exams are quite carefully vetted by both experienced TAs and another very experienced instructor in the department. The physics 1 exam consisted of 15 multiple choice and 5 open-response questions and the physics 2 exam consisted of 20 multiple-choice questions and 6 short-open-response questions. We coded the course exam questions according to whether or not they substantially involved a topic that was listed in the goals (above) of one or more of the lab experiments. On five of the 20 (all open-response) final exam questions for physics 1, the primary concept involved in the question was a concept that was listed in one of the lab goals above. On two of the questions, the concepts covered in the lab were half of what was involved in solving the question. In contrast, 11 of the 20 final exam multiple-choice questions for physics 2 involved primary concepts that were listed in one of the lab goals above. None of the physics 2 open-response questions involved concepts from the lab goals, so these were excluded from the analysis. We coded the final exam questions independently and obtained over 80% agreement, and after brief discussion to clarify what exactly was done in a couple of the labs, converged on 100% agreement.

We then calculated the averages of the scores on lab-related and non-lab-related exam questions for each of the two population groups on each exam. The scores were from the original grading in the course which was done by TAs using rubrics and with considerable care taken to ensure consistency. The grading was done blind to whether or not students were enrolled in the lab. The two split questions for physics 1 were weighted by one half in the calculation of both averages. Finally, we took the ratio (average score on lab-related questions)/(average score on non-lab related questions) for each student and found the average ratio of the averages for the two populations, as shown in table I (physics 1) and table II (physics 2).

The hypothesis was that the lab should improve a student's understanding of content covered in the lab, and this improvement should be reflected in a higher score on exam questions involving those concepts. Hence, those students who took the lab should have a higher ratio than the students who did not. If the lab had no added value, then the ratio for the students who took the lab should be the same as the ratio for the students who did not take the lab. This will be true even if the populations are different.

## IV. Data

The results for the final exams are shown in table I and II. Note that scores shown for the physics 2 exam are higher than that for the physics 1 exam. We attribute this to the use of multiple-choice questions on the physics 2 exam compared to open-response questions on the physics 1 exam.

As has been consistently seen over the years, the students who took the lab are, in general, scoring higher on the overall exam than the students who do not take the lab: physics 2, $t(236.33)=4.86$, $p<.001$. We attribute this to the self-selection of students to take the optional lab course in addition to the lecture course. However, when we compare the ratios of their performance on the lab-related and non-lab-related items, it is clear that the difference between the students that take the lab and those that do not is much smaller than the uncertainty in both cases.

TABLE I. Number of lab-related and non-lab-related questions on the physics 1 final exam and average scores on those questions for students who took the lab and students who did not take the lab. The lab related questions involve 5 + 2 x .5 items, and the non-lab-related are based on 13 + 2 x .5 items. Standard deviations for average scores on both lab and non-lab related questions are all about 25% of totals, unless otherwise specified. The average of all the individual student ratios is shown for the two populations.

| PHYSICS 1 | Number of questions | Score of students who took the lab (N=211) | Score of students who did not take the lab (N=360) |
|---|---|---|---|
| Lab related questions | 6 | $M$=21.83 | $M$=19.00 |
| Non-lab related questions | 14 | $M$=47.24 | $M$=41.57 |
| Average of ratios | | $M$=0.467, $SD$ = 0.10 | $M$=0.468, $SD$ = 0.14 |
| Difference between ratios | 0.001 ± .01; $t$(569)=0.09, $p$= .928 | | |

TABLE II. Number of lab-related and non-lab-related questions on the physics 2 final exam and average scores on those questions for students who took the lab and students who did not take the lab. The lab related questions involve 11 questions, and the non-lab-related are based on the remaining 9 questions. Standard deviations for average scores on both lab and non-lab related questions are all about 20% of totals, unless otherwise specified. The average of all the individual student ratios is shown for the two populations.

| PHYSICS 2 | Number of questions | Score of students who took the lab (N=211) | Score of students who did not take the lab (N=360) |
|---|---|---|---|
| Overall score | 20 | $M$=73.99, $SD$=15 | $M$=66.34, $SD$=16 |
| Lab related questions | 11 | $M$=77.80 | $M$=70.08 |
| Non-lab related questions | 9 | $M$=69.33 | $M$=61.77 |
| Average of ratios | | $M$=1.24, $SD$=0.69 | $M$=1.25, $SD$=0.60 |
| Difference between ratios | -0.01 ± 0.07; $t$(201.81)=-0.15, $p$= .883 | | |

Although we believe that the scores on the final exam are the most relevant, as this was the most inclusive measure and reflects the extent of learning at the end of the course, we also carried out similar analyses for the two midterm exams in physics 1. There were a total of 21 questions, half of which were lab-related. We found similar results; the ratio of scores was slightly higher for students who took the lab on one midterm and slightly lower on the second midterm, but in both cases the differences between lab takers and non-takers were small compared to the standard errors. So, just as on the final exams, there was no measureable impact of taking the lab on the midterm exam scores.

Since many teaching methods have been shown to improve student performance on conceptual questions, but not on more traditional quantitative problem-solving items,[31] we wanted to also explore the type of question included in our analyses. For the physics 2 exam, we coded the lab-related items as to whether they were primarily conceptual or primarily quantitative. We only used the physics 2 exam, since this involved multiple-choice questions. The multi-part structure to the open-response items on the physics 1 exam, as well as their subjective grading by different individuals, made these harder to code and less reliable to examine on a question-by-question basis. The question-by-question analyses for the physics 2 questions are found in table III. Note that since these items were multiple-choice, we provide the percent of students that got the item correct in each group.

Four of the 11 lab-related questions were coded as being primarily conceptual. With a Bonferroni correction applied to account for the multiple comparisons ($\alpha$= .003), only one of the items showed a statistically significant difference between the two groups of students. While this result is statistically significant, we do not feel that modest statistical significance on a single question out of 20 is practically significant. If we combine the four primarily conceptual items and set up a ratio similar to that used in the previous analysis (ratio of conceptual lab-items/other items), again, there is no significant difference between the two groups: $t(311.22)$=-0.09, $p$=.928. This demonstrates that the labs did not measurably affect conceptual learning.

Ratios inherently have non-normal statistical distributions, but in this case the range of values was small compared to the respective means of the ratios, so the distributions looked normal. A more statistically rigorous, but less intuitive to describe, approach is to calculate the percent correct score for each of the groups on the lab-related questions, and take the difference. Then calculate the percent correct score for each of the two groups on the non-related- questions, and take that difference and then examine the difference of the two differences. This calculates an equivalent quantity to the comparison of the ratios, although we found it harder to explain. However, the statistics for this difference of differences in the percent scores are very straightforward as all quantities are linear and have normal distributions. The results, however, are the same as stated above, in that the difference representing the impact of taking the lab course on exam scores is much less than the uncertainty in that difference, and hence consistent with zero.

## V. Discussion and conclusion

Within uncertainties in the differences between the respective ratios, which are about 2% and 1% respectively, of the ratios themselves for Physics I and Physics II, there is no effect on the final exam performance whether or not a student completes the lab course in two different undergraduate physics courses. This is somewhat unexpected, particularly given the high precision of the measurement due to the large sample size and the apparently optimal design and implementation of the lab to maximize learning of the physics content in both of the two courses. It indicates that, relative to the other means by which students learn content in a physics course, such as lectures (particularly the interactive lectures used in these cases), homework, recitation sections, and studying for exams, labs contribute very little.

TABLE III. Question-by-question analysis on the multiple-choice items on the physics 2 final exam and the percentage of students who got it correct from the group of students who did and did not take the lab. Questions were coded as being related (LR) or not related to the lab (NLR). Lab-related items were also coded as being primarily conceptual (CONC) or primarily calculational (CALC). A Bonferroni correction was used to account for the multiple comparisons ($\alpha$= .003).

| Item | Lab-related (LR) or non-lab related (NLR) | Primarily conceptual (CONC) or calculational (CALC) | Percent correct took-lab (N=129) | Percent correct not-take-lab (N=361) | Significant differences |
|---|---|---|---|---|---|
| 1 | NLR | | 91.47 | 90.58 | |
| 2 | NLR | | 85.27 | 77.29 | |
| 3 | LR | CALC | 87.60 | 83.66 | |
| 4 | LR | CALC | 67.44 | 61.50 | |
| 5 | LR | CALC | 72.09 | 64.82 | |
| 6 | LR | CALC | 93.02 | 87.26 | |
| 7 | NLR | | 83.72 | 80.33 | |
| 8 | NLR | | 86.05 | 75.90 | |
| 9 | NLR | | 65.12 | 59.00 | |
| 10 | LR | CONC | 46.51 | 39.89 | |
| 11 | LR | CALC | 58.91 | 45.15 | |
| 12 | LR | CONC | 61.24 | 60.94 | |
| 13 | LR | CONC | 94.57 | 86.70 | |
| 14 | NLR | | 55.04 | 42.94 | |
| 15 | NLR | | 58.91 | 46.26 | |
| 16 | LR | CONC | 87.60 | 72.30 | $z$=3.51, $p$=.0005 |
| 17 | LR | CALC | 93.80 | 84.49 | |
| 18 | NLR | | 35.66 | 29.64 | |
| 19 | NLR | | 62.79 | 54.02 | |
| 20 | LR | CALC | 93.02 | 84.21 | |

There are some caveats to this result. These exam questions were not written to specifically address what was covered in the labs, only to test students on their mastery of the material covered in the regular course. The assumption underlying our methodology is that if an exam problem involves correctly applying a concept or procedure covered by a lab experiment, students will perform better on that question if the lab resulted in their having a better understanding of that material. This may not be the case. As previously mentioned, it has been demonstrated that particular teaching methods result in improved scores on concept inventories but have negligible effect on more traditional quantitative test questions.[31] Examining primarily conceptual questions only still did not demonstrate the added-value of taking the lab, and many of the labs also involved calculations similar to those required on the exams.

This study does not rule out the possibility that there are other things learned in the lab that are not being tested by the course exams. Indeed, the AAPT has recently published an updated set of goals for undergraduate physics lab curriculum,[32] which are much more skills-focused. These goals fall under the following themes: analyzing and visualizing data, communicating physics, constructing knowledge, designing experiments, developing technical and practical skills; and modeling. A striking difference between this report and the 1998 report[14] is the removal of conceptual learning as an outcome. It is possible that students were learning some of the skills listed in the AAPT document, even though they were not the stated goals of these individual labs.

While these caveats offer possible reasons that the impact of the lab course might not have been visible in the course exam scores, the results still should raise concern. Given the resources required for instructional laboratories and the amount of student time invested in them, one would hope it should be quite easy to observe their educational impact. We hope this work will inspire many institutions to examine carefully the learning objectives of their introductory lab courses, and the learning outcomes actually being achieved by those courses. This work indicates that, in the absence of evidence, it would be a mistake to assume that all that time and money is being well spent.

## Acknowledgements

We are pleased to acknowledge the assistance of Chaya Nanavati and the instructors of the respective physics courses and lab courses in carrying out this work.

---

## A. Goals for physics lab 1

1. *"The goals of this lab are two-fold:*
   - *To introduce you to the track-cart Logger Pro system that you will be using for a few labs.*
   - *To give you a better understanding of the mathematical relationships between position and time, velocity and time, and acceleration and time (see Chapters 1 and 2 in Knight.)*
2. *The goal of this lab is to give you a better understanding of the physical meaning of graphs depicting force-versus-time (force plots) so you can describe the physical "story" behind a force plot.*
3. *The goals of this lab are to understand*
   - *The vector nature of forces;*
   - *The interaction of multiple forces in two dimensions.*
4. *The goal of this lab is to give you a better understanding of free body diagrams and motion in the presence of friction.*
5. *Activity I: The goal of this activity is to gain a better understanding of frames of reference and relative motion.*
   *Activity II: The goal of this lab is to literally "feel" the central inward force associated with circular motion and to understand how the mass, speed, and radius of the circular motion affect the force.*
6. *The purpose of this lab is to give you some hands-on exposure to collisions on an idealized, almost-frictionless track. (Knight, 9.4-9.6)*
7. *Energy is neither created nor destroyed but can be converted from one form to another. In this lab you will explore two different forms of energy, kinetic and potential. (Knight sections 10.2, 10.4, 10.5)*
8. *This lab will allow you to review concepts that you have been exposed to in [physics 1], specifically the physics of projectiles and the conservation laws."*
9. Rigid body dynamics. No goals were stated, but the experiments all involved looking at rotation of rigid bodies, and exploring and making sense of the relationship between linear and angular acceleration and between force and toque.

## B. Goals for physics lab 2

1. *"The purpose of this lab is to*
   - *Understand how to select appropriate graphs to understand physical relationships*
   - *And to use this knowledge to verify Coulomb's Law*
2. *The goal of the lab is to make you familiar with the equipment and instruments that you will use in future labs -- digital multi-meters (DMMs), breadboards, function generators, and oscilloscopes.*
3. *The purpose of this lab is to*
   - *Understand how charges distribute on a conductor*
   - *Map electric potentials for various charge configurations and*
   - *Develop an intuitive understanding of the relationship between fields and potentials*
4. *The goal of this lab is to gain familiarity with charging/discharging capacitors and with calculating and measuring the associated time constants, and to understand and measure* the effective capacitance when two or more capacitors are connected in series or parallel.

5. *The purpose of this lab is to give you a better understanding of basic circuits with resistors and/or capacitors. You will then use your understanding to design circuits for specific functions.*

6. *This lab has two independent parts that you can complete in any order.*
   - *In the first part, you will use a magnetic field produced by a pair of current-carrying coils to deflect an electron beam and measure the charge to mass ratio for an electron.*
   - *In one part, you will use the magnetic field of a pair of current-carrying coils to measure the earth's magnetic field.*

7. *The goals of this lab are: To understand the concept of magnetic flux and to understand what happens when the magnetic flux through a wire loop is changed over time*

8. *The goal of this lab is to*
   - *Clarify the concept of "ground" in a circuit*
   - *Learn about time-dependent LR (or RL) circuits. The main emphasis is on studying the response of these circuits to step changes in voltage e.g. by flipping a switch (open or closed) and exposing the circuit elements to a constant voltage.*
   - *If there is time left, your TA will suggest some experiments with sinusoidally varying voltages and LR circuits.*

9. *As part of this lab you will:*
   - *Continue to work with breadboards, oscilloscopes, and grounds*
   - *Obtain a deeper understanding of the material presented in Chapter 35 of Knight*
   - *Specifically, you should have a better understanding of AC circuits"*